\documentclass[11pt,a4paper]{article}
\usepackage{microtype}
 \usepackage{ifpdf}
 \ifpdf
 \usepackage[pdftex]{graphicx}
 \else
\usepackage{graphicx}
 \fi
 \usepackage{caption}
 \usepackage{subcaption}
\usepackage{booktabs}
\usepackage[T1]{fontenc}
\usepackage{textcomp}
\usepackage{float}
\usepackage[sc,noBBpl]{mathpazo}
\usepackage{amsmath,amsfonts,amssymb,amsthm}
\usepackage[mathscr]{eucal}
\usepackage[square,numbers]{natbib}

\newcommand\scN{{\mathscr N}}

\newcommand\mvector{\boldsymbol}

\newcommand\vd{\mvector{d}}

\newcommand\vp{\mvector{p}}
\newcommand\vq{\mvector{q}}

\newcommand\vGamma{\mvector{\Gamma}}

\newcommand\field{\mathbb}

\newcommand\R{\field{R}}
\newcommand\bbS{\mathbb{S}}
\newcommand\EE{\field{E}}
\newcommand\C{\field{C}}
\newcommand\Z{\field{Z}}

\newcommand\bbH{\field{H}}

\newcommand\rmd{\mathrm{d}}

\newcommand\rmi{\mathrm{i}\mspace{1mu}}

\newcommand\abs[1]{\lvert #1 \rvert}

\newcommand\sk{\operatorname{S}_{\kappa}}
\newcommand\ck{\operatorname{C}_{\kappa}}

\newcommand\defset[2]{\left\{{#1}\;\vert \;\; {#2} \,\right\}}

\usepackage[square]{natbib}
\usepackage[square]{natbib}
\theoremstyle{plain}
\newtheorem{theorem}{Theorem}
\newtheorem{lemma}[theorem]{Lemma}

\newtheoremstyle{note}{\topsep}{\topsep}{\slshape}{}{\scshape}{}{ }{}
\theoremstyle{note}

\newtheorem{remark}[theorem]{Remark}
\numberwithin{equation}{section}
\numberwithin{theorem}{section}
\title{Note on integrability of certain homogeneous Hamiltonian systems in 2D constant curvature spaces}
\author{
    Andrzej J.~Maciejewski$^1$, Wojciech Szumi\'nski$^2$,  \\ 
   and  Maria Przybylska$^2$ \\[1em]
  {}$^{1}$Janusz Gil Institute of  Astronomy, \\ University of Zielona G\'ora, 
  Licealna 9,  \\
  PL-65-407,  Zielona G\'ora, Poland\\
e-mail: maciejka@astro.ia.uz.zgora.pl \\[1em]
   {}$^2$ Institute of Physics,\\ University of Zielona G\'ora, 
  Licealna 9,  \\
  PL-65-407,  Zielona G\'ora, Poland\\ e-mail: w.szuminski@if.uz.zgora.pl \\
   e-mail: M.Przybylska@if.uz.zgora.pl }
\begin{document}
%
\maketitle
\begin{abstract}
We formulate the necessary
  conditions for the integrability of a certain family of
  Hamiltonian systems defined in the constant curvature two-dimensional
  spaces. Proposed form of potential can be considered as a counterpart of a homogeneous potential in flat spaces.
 Thanks to this property Hamilton equations admit, in
a general case, a particular solution. Using this solution we derive necessary
integrability conditions  investigating
differential Galois group of variational equations.   
\end{abstract}
\date{\small Key words: integrability obstructions; Liouville integrability; Morales--Ramis theory; differential Galois theory;   constant curvature spaces }
\maketitle

%
\section{Introduction}
Integrability  of natural Hamiltonian systems of the form
\begin{equation}
  \label{eq:1}
  H = \frac{1}{2}\sum_{i=1}^n p_i^2 +V (\vq), \qquad \vq=(q_1, \ldots,q_n)
\end{equation}
has been  intensively investigated during last decades and 
significant successes were achieved. Here $\vq=(q_1, \ldots,q_n)$ and  $\vp=(p_1, \ldots, p_n)$ are canonical
variables in $\C^{2n}$ considered as a symplectic linear space.
It seems that among new methods which have been
invented, the most powerful and efficient are those formulated in the
framework of the differential Galois theory. The necessary conditions
for the integrability of a Hamiltonian system in the Liouville sense
are given in terms of properties of the differential Galois group of
variational equations obtained by linearisation of equations of
motion in a neighbourhood of a particular solution. The fundamental
Morales-Ramis theorem of this approach says that if a Hamiltonian
system is meromorphically integrable in the Liouville sense in a
neighbourhood of a phase curve $\vGamma$ corresponding to a particular
solution, then the identity component $\mathcal{G}^0$ of the
differential Galois group $\mathcal{G}$ of variational equations along
$\vGamma$ is Abelian, see e.g. \cite{Morales:01::a,Morales:99::c}.

To apply the above mentioned method we have to know a
particular solution of the system but, in general, we do not know how
to find it. Fortunately, for Hamiltonian systems~\eqref{eq:1} with the
potential $V(\vq)$ which is homogeneous of degree $k\in\Z$, we know
that in a generic case they admit  particular solutions of the form
\begin{equation}
  \label{eq:2}
  \vq(t) = \varphi(t) \vd, \qquad \vp(t)= \varphi(t) \vd, \qquad \ddot
  \varphi = -\varphi ^{k-1}.
\end{equation} 
where $\vd\in\C^n$ is a non-zero solution of the non-linear system
$V'(\vd)=\vd$.  Moreover, variational equations along these particular
solutions can be transformed into a system of uncoupling
hypergeometric equations depending on the degree of homogeneity $k$
and eigenvalues $\lambda_i$, for $i=1,\ldots,n$, of the Hessian
$V''(\vd)$. Since differential Galois group of the hypergeometric
equation is well known it was possible to obtain necessary conditions of
the integrability of Hamiltonian systems \eqref{eq:1} in the form of
arithmetic restrictions on $\lambda_i$ that must belong to appropriate
sets of rational numbers depending on $k$, see
e.g.~\cite{Morales:99::c,Morales:01::a}. Later it appeared that
between $\lambda_i$ some universal relations exist which improves
conditions mentioned in the above papers, for details see e.g. \citep{Maciejewski:05::b,mp:09::a,Maciejewski:09::}.

Successful integrability analysis of Hamiltonian systems with
homogeneous potentials in flat Euclidean spaces motivated us  to  look for 
systems in curved spaces with similar properties. For natural systems
\begin{equation}
\label{eq:3}
H= \frac{1}{2} \sum_{i,j=1}^n g^{ij}p_ip_j + V(\vq)
\end{equation}
defined on $T^{\star}M^n$, where $M^{n}$ is a Riemannian manifold with metric $g=\{g_{ij}\}$, there is  a good notion of homogeneous functions. In
paper \cite{mp:15::d} we proposed to study the following form
of the Hamiltonian
\begin{equation}
  \label{eq:m_h}
  H=T+V,\qquad
  T=\frac{1}{2}r^{m-k}\left(p_r^2+\frac{p_\varphi^2}{r^2}\right),
  \qquad	 V=r^m U(\varphi),
\end{equation}
where $m$ and $k$ are integers, $k\neq 0$ and $U(\varphi)$ is a
meromorphic function.  If we consider $(r,\varphi)$ as the polar
coordinates, then the kinetic energy corresponds to a flat singular
metric on a plane.  This is just an example of a natural system which
possesses certain common features with standard Hamiltonian systems with
homogeneous potentials in the Euclidean plane $\EE^2$.

In this paper we propose another class of natural Hamiltonian systems
with two degrees of freedom defined on $T^{\star}M^2$ where $M^2$ is a two
dimensional manifold with a constant curvature metrics. More
specifically, $M^2$ is either sphere $\bbS^2$, the Euclidean plane $\EE^2$,
or the hyperbolic plane $\bbH^2$ with curvature parameter $\kappa$ positive, equal to zero or negative, respectively.
In order to consider those three
cases simultaneously we will proceed as in
\cite{Herranz:00::,Ranada:99::a} and we define the following $\kappa$-dependent trigonometric functions
\begin{equation}
  \label{eq:ck}
  \ck(x):= 
  \begin{cases} 
    \cos(\sqrt{\kappa}x) &\text{for} \quad\kappa>0 \\
    1 & \text{for}\quad  \kappa=0 \\
    \cosh(\sqrt{-\kappa}x)&\text{for} \quad \kappa<0
  \end{cases}
,\quad
  \sk(x):= 
  \begin{cases} 
    \frac{1}{\sqrt{\kappa}}\sin(\sqrt{\kappa}x) &\text{for} \quad\kappa>0 \\
    x & \text{for} \quad\kappa=0 \\
    \frac{1}{\sqrt{-\kappa}}\sinh(\sqrt{-\kappa}x)&\text{for}\quad
    \kappa<0
  \end{cases}.
\end{equation}
These functions satisfy the following identities
\begin{equation}
  \ck^2(x)+\kappa \sk^2(x)=1,\quad \sk'(x)=\ck(x),\quad \ck'(x)=-\kappa\sk(x).
  \label{eq:trigono}
\end{equation}
Our aim is to study  the following  natural Hamiltonian systems 
\begin{equation}
  H=T+V(r, \varphi),\qquad T=\frac{1}{2}\left(p_r^2+\dfrac{p_{\varphi}^2}{\sk(r)^2}\right),\quad V(r, \varphi)=\sk^m(r)U(\varphi),
  \label{eq:homo}
\end{equation}  
where $m\in\Z$ and $U(\varphi)$ is a  meromorphic function of
variable $\varphi$. This is a natural Hamiltonian system defined
on $T^{\star}M^2$ for the prescribed $M^2$. Notice that the kinetic
energy as well as the potential depends on the curvature $\kappa$. It
appears that for such Hamiltonian systems we can find certain
particular solutions and we are able to perform successfully
differential Galois integrability analysis. 

To apply the Morales-Ramis theory we consider the complex version of our system. We assume that there exists   $\varphi_0\in\C$ such that $U'(\varphi_0)=0$ and
  $U(\varphi_0)\neq 0$. Under these assumptions we define
  \begin{equation}
    \label{eq:4}
    \lambda := 1 + \frac{U''(\varphi_0)}{m U(\varphi_0)}. 
  \end{equation} 
The main result of this
paper is the following theorem that gives necessary conditions for the
integrability of Hamiltonian systems governed by
Hamiltonian~\eqref{eq:homo}.   
\begin{theorem}
  If the Hamiltonian system governed by Hamilton function
  \eqref{eq:homo} with $m\kappa\neq 0$ is meromorphically integrable,
  then the pair $(m,\lambda)$ belongs to the following list
  \begin{equation}
    \label{eq:tab}
    \begin{array}{cll}
      \text{case} &  m& \lambda \\[0.5em]
      1&m&-\dfrac{(m - 2 p) (p-1)}{m}\\[0.9em]
      2&m&-\dfrac{(m + 4 p) [m - 4 (1 + p)]}{8 m}\\[0.9em]
      3&-2+4p& \text{arbitrary}\\
      4&m=2q-1&  -\dfrac{(-2 + 3 m + 12 p) [3 m - 2 (5 + 6 p)]}{72 m}
    \end{array}
  \end{equation}
  Here $p$ and $q$ are  arbitrary integers.
  \label{th:1}
\end{theorem}
Case $\kappa=0$ is excluded from the above theorem because it is
already covered by Theorem~5.1 in \citep{Morales:99::c,Morales:01::a}. Case $m=0$
and $\kappa=0$ was already considered in \citep{mp:10::a} while case
$m=0$ and $\kappa\neq 0$ needs  separate investigations.

Let us notice that conditions for integrability given in the above
theorem neither depend on the sign, nor on the value of the
curvature. This fact is related to the homogeneity of the system. Using real
scaling we can reduce the curvature to three values
$\kappa\in\{-1,0,1\}$. As we work in complex variables we can always
choose variables and time such that $\kappa$ is either zero or one.
\section{Proof of Theorem~\ref{th:1}}
\label{sec:proof}
Hamilton equations governed by Hamiltonian \eqref{eq:homo} have the form
\begin{equation}
 \begin{split}
&\dot r=p_r,\qquad\quad \dot p_r= \dfrac{p_{\varphi}^2}{\sk^3(r)} \ck(r)-m\sk^{m-1}(r)\ck(r)U(\varphi),\\
&\dot \varphi= \dfrac{p_{\varphi}}{\sk^2(r)},\,\quad \dot p_{\varphi}=-\sk^{m}(r)U'(\varphi).
 \end{split}
 \label{eq:m_vh}
\end{equation} 
If $U'(\varphi_0)=0$ for a certain $\varphi_0\in\C$, then
system~\eqref{eq:m_vh} has a two dimensional invariant manifold of the form
\begin{equation}
  \label{eq:vh0_N}
  \scN=\left\{(r, p_r, \varphi,p_\varphi )\in \C^4\ |\ \varphi=\varphi_0, \     p_\varphi=0\right\}.
\end{equation}
Invariant manifold $\scN$  is foliated by phase curves parametrised by energy $e$
\[
 e=\frac{1}{2}p_r^2+\sk^m(r)U(\varphi_0),
\]
that gives us particular solutions. If $[R,\Phi,P_R,P_{\Phi}]^T$ denote the variations of variables $[r,\varphi,p_r,p_{\varphi}]^T$, then variational equations along a particular  solution lying on $\scN$ take the  form
\begin{equation}
 \begin{bmatrix}
  \dot R\\
\dot \Phi\\
\dot P_R\\
\dot P_{\Phi}
 \end{bmatrix}=
\begin{bmatrix}
 0&0&1&0\\
0&0&0&\sk^{-2}(r)\\
m\sk^{m-2}(r)[m\kappa \sk^2(r)-(m-1)]U(\varphi_0)&0&0&0\\
0&-\sk^m(r)U''(\varphi_0)&0&0
\end{bmatrix}
\begin{bmatrix}
  R\\
 \Phi\\
 P_R\\
 P_{\Phi}
 \end{bmatrix}.
 \label{eq:variaty}
\end{equation}   
Since the motion takes place in the plane $(r,p_r)$ the normal part of variational equations is given by the following closed subsystem
\[
 \dot \Phi=\sk^{-2}(r)P_{\Phi},\qquad \dot P_{\Phi}=-\sk^m(r)U''(\varphi_0)\Phi,
 \]
 or rewritten as a one second order equation
 \[
\ddot \Phi+a(r,p_r)\dot \Phi+b(r,p_r)\Phi=0,\qquad a(r,p_r)=2\frac{\ck(r)}{\sk(r)}p_r,\quad b(r,p_r)=\sk^{m-2}(r)U''(\varphi_0).
\]
Using the change of independent variable $t\mapsto z=\sk(r)/\sqrt{\kappa}$  we can transform it into a  linear equation with rational coefficients
\begin{equation}
\begin{split}
& \Phi''+p(z)\Phi'+q(z)\Phi=0,\\
&p(z)=\dfrac{\ddot z +\dot z a}{\dot z^2}=  \frac{z}{z^2-1}+2\frac{(k+2)z^m -u^m}{z(z^m-u^m)} 
,\\
&q(z)=\dfrac{b}{\dot z^2}=\dfrac{m ( \lambda-1) z^{m-2}}{2 (z^2-1) (z^m -u^m)},
\end{split}
\label{eq:pierwsze}
\end{equation} 
where $u$ is defined by relation $ e\kappa^{m/2}=U(\varphi_0)u^m$.
Notice that $u$ is a singular point of the equation which depends on
the choice of the energy. In order to analyse effectively  cases with arbitrary $m$ we fix $u=0$. It is equivalent to take zero energy level. The variational equation~\eqref{eq:pierwsze} reduces to the following one
\[
\Phi''  +\dfrac{(m+6) z^2 -4 - m }{2 z ( z^2-1)}\Phi'' +\dfrac{k (\lambda-1 )}{2 z^2 ( z^2-1)}\Phi=0.
\]
This equation has three regular singular points at $z\in\{-1,0,1\}$. 
Using transformation $z\mapsto y=z^2$ we transform it into equation 
\begin{equation}
 \dfrac{\mathrm{d}^2\Phi}{\mathrm{d} y^2}+\left[\frac{m+6}{4 y}+\frac{1}{2 (y-1)}\right]
\dfrac{\mathrm{d}\Phi}{\mathrm{d} y}
+\left[
\dfrac{m(1 - \lambda)}{8 y^2}+\dfrac{m (\lambda-1)}{8 y(y-1)}\right]\Phi=0,
\label{eq:norm}
\end{equation}
which is the Riemann $P$ equation
\begin{equation}
\label{eq:riemann1}
\begin{split} 
\dfrac{\mathrm{d}^2\Phi}{\mathrm{d}y^2}&+\left[\dfrac{1-\alpha-\alpha'}{y}+ 
\dfrac{1-\gamma-\gamma'}{y-1}\right]\dfrac{\mathrm{d}\Phi}{\mathrm{d}y}+ 
\left[\dfrac{\alpha\alpha'}{y^2}+\dfrac{\gamma\gamma'}{(y-1)^2}+ 
\dfrac{\beta\beta'-\alpha\alpha'-\gamma\gamma'}{y(y-1)}\right]\Phi=0, 
\end{split} 
\end{equation}
 see e.g.~\cite{Whittaker:35::}.
For the considered equation  the respective  differences of exponents at singularities $y=0$, $y=1$ and $y=\infty$ are the  following
\begin{equation}
 \rho=\alpha-\alpha'=\frac{1}{4} \sqrt{( m-2)^2 + 8 m \lambda},\quad
\tau=\gamma-\gamma=\frac{1}{2},\quad\sigma=\beta-\beta'=\frac{m+4}{4}. 
\label{eq:difery}
\end{equation}
If the system is integrable, then according to the Morales-Ramis
theorem, the identity components of the differential Galois group of
variational equations \eqref{eq:variaty} as well as of normal variational
equation~\eqref{eq:norm} are Abelian. If the identity component of the
differential Galois group is Abelian, then in particular it is
solvable.  But necessary and sufficient conditions for solvability of
the identity component of the differential Galois group for the
Riemann $P$ equation are well known and formulated in the Kimura
theorem which we recall in Appendix~\ref{sec:hyper}, for details see
\cite{Kimura:69::}. These conditions are expressed in terms of
conditions on differences of exponents~\eqref{eq:difery}.  The proof of
Theorem~\ref{th:1} consists in a direct application of
Theorem~\ref{th:Kimura} to our Riemman P equation~\eqref{eq:norm}
and taking into account that $m$ has to be an integer.

The condition A of Theorem~\ref{th:Kimura} is fulfilled if at least
one of  the numbers  $ \rho+ \tau+ \sigma$, $-  \rho+ \tau+ \sigma$, $  \rho- \tau+ \sigma$, $  \rho+ \tau- \sigma$ is
 an odd integer. This condition gives  the following forms for $\lambda$
\[
 \lambda=-\dfrac{(m - 4 p) (2 p-1)}{m},\quad\text{or}\quad \lambda=\dfrac{2 p (  m-2 + 4 p)}{m},
\]
where $p\in\Z$. Let us notice that these two formulae can be obtained from the following one
\[
 \lambda=-\dfrac{(m - 2 p) (p-1)}{m},
\]
for $p$ even and odd, respectively.

In Case B of Theorem~\ref{th:Kimura} the
quantities $\rho$ or $-\rho$, $\sigma$ or $-\sigma$ and $\tau$ or
$-\tau$ must belong to Table~\ref{tab:sch_app} that is called Schwarz's table.
As $\sigma=\tfrac{1}{2}$ only items $1$, $2$, $4$, $6$, $9$, or $14$
of Table~\ref{tab:sch_app} are allowed and we will analyse them case by case. In  calculations below numbers $p$ and $q$ are integers.

In item $1$ the choice $\pm\rho=1/2+p$ and arbitrary  $\sigma$  gives
\begin{equation}
 \lambda=-\dfrac{(m + 4 p) [m - 4 (1 + p)]}{8 m},
\label{eq:lam1/2}
\end{equation}
and no obstructions on $m$.
The second possibility in this item that $\pm\sigma=1/2+p$ and  $\rho$ is arbitrary
gives $m=-2+4p$ and no obstructions for $\lambda$.

In  item $2$, condition $\pm\rho=1/2+p$ gives \eqref{eq:lam1/2} and 
$\sigma=1/3+q$ leads 
to non-integer $m=-8/3 + 4 q$.

In item $4$ the choice $\pm\sigma=1/3+q$ leads to non-integers $m=-8/3 + 4 q$ and $m=-16/3 - 4 q$.
The second possibility in this item:  $\pm\rho=1/3+p$ gives 
\[
\lambda= -\dfrac{(-2 + 3 m + 12 p) [3 m - 2 (5 + 6 p)]}{72 m}
\]
and conditions  $\pm\sigma=1/4+q$ lead to  $m=-3 + 4 q$ or $m=-5 - 4 q$, respectively. Elements of both these sets can be written as $m=2q-1$.

In item $6$ the choice $\pm\sigma=1/3+q$ leads to non-integers $m=-8/3 + 4 q$ or $m=-16/3 - 4 q$.
Also  the second possibility $\pm\sigma=1/5+q$ gives only non-integer $m=-16/5 + 4 q$ or $m=-24/5 - 4 q$.

Similarly in item $9$ both possibilities  lead to non-integer $m$. Namely, choice $\pm\sigma=1/5+q$ gives
non-integer $m=-16/5 + 4 q$ or $m=-24/5 - 4 q$ and 
 $\pm\sigma=2/5+q$  produces only non-integer  $m=-12/5 + 4 q$ or $m=-28/5 - 4 q$.

In item $14$ the choice $\pm\sigma=2/5+q$ gives non-integers $m=-12/5 + 4 q$ or
$m=-28/5 - 4 q$, respectively. Similarly the second possibility in this item that $\pm\sigma=1/3+q$ leads only to non-integers $m=-8/3 + 4 q$ or $m=-16/3 - 4 q$.

Collecting all admissible forms for $\lambda$ and $m$ we obtain list in Eq.~\eqref{eq:tab}.
\section{Remarks and examples}

Result of Theorem~\ref{th:1} is not optimal. In the proof we selected
a phase curve with zero energy. We did this because for this energy
value the normal variational equation reduces to the Gauss
hypergeometric equation for arbitrary $m$.  For other energy values normal variational equation remains Fuchsian but the number of singularities depends on
$m$ and we did not know how to analyse properties of its differential
Galois group for an arbitrary $m$.  For small values of $\abs{m}$, one
can apply the Kovacic algorithm to test if the identity component of
the differential Galois group is solvable.  In this way we can show
the following.
\begin{lemma}
  If Hamiltonian system given by \eqref{eq:2} with $\kappa\neq 0$ and
  $m=1$ is integrable, then either $\lambda=1$, or $\lambda=0$.
\end{lemma}
However, for $\abs{m}>2$ as well as for $m=-1$ there is no an
effective way to perform the Kovacic algorithm till the end because
its splits into infinitely many cases for which we have to determine
if certain systems of algebraic equations have solutions.

If $m=-2$ the Hamiltonian system~\eqref{eq:homo} is integrable. The
additional first integral has the following form
\begin{equation}
  G=\frac{p_\varphi^2}{2}+U(\varphi).
\end{equation}
In fact, in this case system is separable in variables
$(r,\varphi)$. For such systems one can ask about its
superintegrability and in \cite{mp:10::c} it was shown that the
necessary condition is that $\lambda=1-s^2$, where $s$ is a non-zero
rational number.

It is known, see e.g. \citep{Morales:99::c,Morales:01::a}, that for
$m=2$ and $\kappa=0$ we cannot deduce any obstruction for the
integrability from analysis of variational equations for prescribed
straight-line solutions. Our theorem for $m=2$ also does not give
obstructions for integrability because it belongs to case 3 in Table
\eqref{eq:tab}. However, one can suspect that our weak conclusion is
an effect of considering a peculiar energy level. But it is not this case.
For $m=2$ variational equation~\eqref{eq:pierwsze} is always solvable with
solutions
\begin{equation}
  \label{eq:5}
  \Phi_{\pm}(z)=\frac{1}{z} \sqrt{z^2+\lambda-1}\exp \left[ \pm \int \omega(z) \rmd z \right],
\end{equation}
where
\begin{equation}
  \label{eq:6}
  \omega(z)^2 = \frac{\lambda(1-\lambda)(u^2+\lambda-1)}{(z^2+\lambda-1)^2[z^4-(1+u^2)z^2+u^2]}.
\end{equation}

As our first example we consider the Hamiltonian~\eqref{eq:homo} with the
potential
\begin{equation}
  \label{eq:pot}
  V(r,\varphi)=S_\kappa^m(r)\cos^m\varphi, 
\end{equation}
for which $\varphi_0=0$ and $\lambda=0$, so the necessary conditions
for integrability are fulfilled. In fact the system is integrable, and
the additional first integral is
\begin{equation}
  \label{eq:int1}
  I_{\kappa}=p_r\sin\varphi +
  p_\varphi\cos\varphi\dfrac{C_\kappa(r)}{S_\kappa(r)}, \qquad \kappa\neq 0.
\end{equation}
We notice that it does not depend on $m$. Limit
\begin{equation}
  \label{eq:8}
  I_0=\lim_{\kappa\rightarrow 0} I_{\kappa} =p_r\sin\varphi +r^{-1}p_\varphi\cos\varphi,
\end{equation}
gives the first integral for the case $\kappa=0$.  If $\kappa=0$ and
$m=1$, then there exists additional independent first integral
quadratic in momenta
\begin{equation}
  I_2=\left(p_r^2-\frac{p_\varphi^2}{r^2}\right)\cos\varphi\sin\varphi+r^{-1}
  p_rp_\varphi\cos(2\varphi)-r\sin
  \varphi.
   \label{eq:int2}
\end{equation}
Thus, in this case the system is maximally super-integrable. Failure
of direct search of additional first integral up to degree three in
momenta for $\kappa\neq0$ is in accordance with results of paper
\cite{Ranada:99::a} where authors gave classification of
superintegrable systems on $\bbS^2$ and $\bbH^2$ with additional first
integrals quadratic in momenta. In particular in this paper it was shown that for our
potential $V=S_\kappa^m(r)\cos^m\varphi$ other terms  must be added in
order to get superintegability with quadratic first integrals.

For the potential
\begin{equation}
  \label{eq:pot1}
  V(r,\varphi)=S_\kappa^m(r)\sin^m\varphi, 
\end{equation}
$\varphi_0=\tfrac{\pi}{2}$ and $\lambda=0$, thus necessary
integrability conditions are satisfied.
We notice that it is obtained from \eqref{eq:pot} by shifting the angle
$\varphi\to \tfrac{\pi}{2}+\varphi$. Thus
 we can immediately write
integrable cases replacing in the above formulae \eqref{eq:int1} -- \eqref{eq:int2} for first integrals
$\sin\varphi\to \cos\varphi$ and $\cos\varphi\to
-\sin\varphi$. Potential \eqref{eq:pot1} for $m=2$ corresponds to the
special case of the so-called anisotropic Higgs oscillator with
$\delta=0$, see Section~3 in \cite{Ballesteros:13::} and Section~5 in
\cite{Ballesteros:14::}.
\begin{remark}

Hamiltonian  systems described in~\eqref{eq:homo} are obtained by a restriction of natural systems in $\R^3$ given by  the following Lagrangian
  
\begin{equation}
\label{eq:11}
L = \frac{1}{2\kappa} \left[ \dot x_0^2+ \kappa \left( \dot x_1^2+ x_2^2 \right)\right]-W(x_0,x_1,x_2), 
\end{equation}
to a quadric
\begin{equation}
\label{eq:12}
\Sigma=\defset{(x_0,x_1,x_2)\in\R^3}{x_0^2+\kappa(x_1^2+x_2^2)=1}.
\end{equation}
Variables $ (r,\varphi)$  are just local coordinates on $\Sigma$.
We have
\[
    x_0=\ck(r),\quad x_1=\sk(r)\cos\varphi,\quad
    x_2=\sk(r)\sin\varphi,
  \]
and  $V(r,\varphi)=W(x_0,x_1,x_2)$.

It is instructive to take $(x_1,x_2)$ as local coordinates on $\Sigma$.
Then the Lagrangian  has the form
 \[
    L= \dfrac{1}{2}\left(\dot
      x_1^2+\dot x_2^2\right)+\dfrac{\kappa(x_1\dot x_1+x_2\dot
      x_2)}{2\left(1-\kappa(x_1^2+x_2^2)\right)} - V(x_1,x_2),
  \]
  where now $V(x_1,x_2)=W(x_0,x_1,x_2)$, with
  $x_0^2=1-\kappa(x_1^2+x_2^2)$. Denoting by $(y_1,y_2)$ the momenta conjugated to $(x_1,x_2)$ we obtain
  \[
    y_1=\dot x_1+\dfrac{\kappa x_1(x_1\dot x_1+x_2\dot
      x_2)}{1-\kappa(x_1^2+x_2^2)},\quad y_2=\dot x_2+\dfrac{\kappa
      x_2(x_1\dot x_1+x_2\dot x_2)}{1-\kappa(x_1^2+x_2^2)}
  \]
  and the Hamilton function of geodesic motion takes the form
  \begin{equation}
    H=\dfrac{1}{2}\left[y_1^2+y_2^2-\kappa(x_1y_1+x_2y_2)^2\right]+
    V(x_1,x_2).
    \label{hamamb}
  \end{equation}
Notice that now the  potential
  \eqref{eq:pot} in ambient coordinates is just $x_1^m$. The first
  integral $I_{\kappa}$ given by \eqref{eq:int1} in ambient variables
  takes the form
  \[
    I_{\kappa}= y_2\sqrt{1-\kappa(x_1^2+x_2^2)}.
  \]
  This function is  a first integral for
  arbitrary potential $V=f(x_1)$.
\end{remark}

\begin{figure}[H]
  \begin{center}
    \resizebox{82mm}{!}{\input{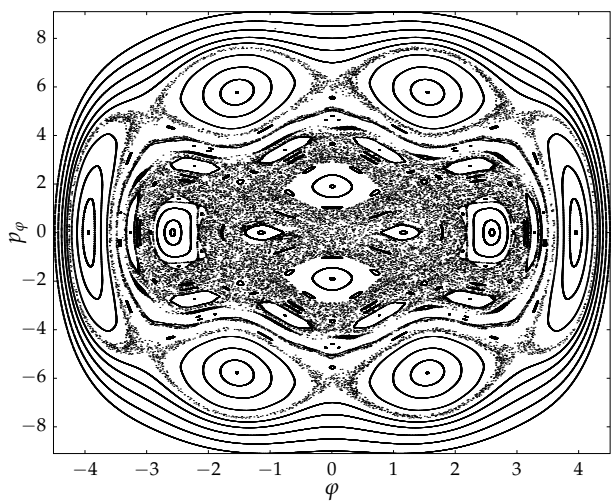}}
    \caption{\small The Poincar\'e cross section for the
      Hamiltonian~\eqref{eq:homo} with potential
      $V=\sin^{-1}r\cosh\varphi$ on energy level $e=50$. The
      cross-section plane is $r=\pi/2$ with $p_r>0$\label{fig:km1}}
  \end{center}
\end{figure}

As the second example we consider the following family of potentials
\begin{equation}
  \label{eq:potential example 3}
  V(r,\varphi)=S_\kappa^m(r)\cosh\varphi.
\end{equation}
\begin{figure*}[h]
  \begin{center}
    \begin{subfigure}[b]{0.47\textwidth}
      \label{fig:angle25a}
      \resizebox{75mm}{!}{\input{poin1.tex}}
      \caption{global regular pattern \label{fig:km3glob}}
    \end{subfigure}\quad\quad
    \begin{subfigure}[b]{0.47\textwidth}
      \resizebox{75mm}{!}{\input{poin2.tex}}
      \caption{magnification of chaotic region \label{fig:km3mag}}
    \end{subfigure}
    \caption{\small The Poincar\'e cross section for the
      Hamiltonian~\eqref{eq:homo} with potential
      $V=\sin^{-3}r\cosh\varphi$ on energy level $e=50$. The
      cross-section plane is $r=\pi/2$ with $p_r>0$\label{fig:km3}}
  \end{center}
\end{figure*}
\begin{figure*}[h]
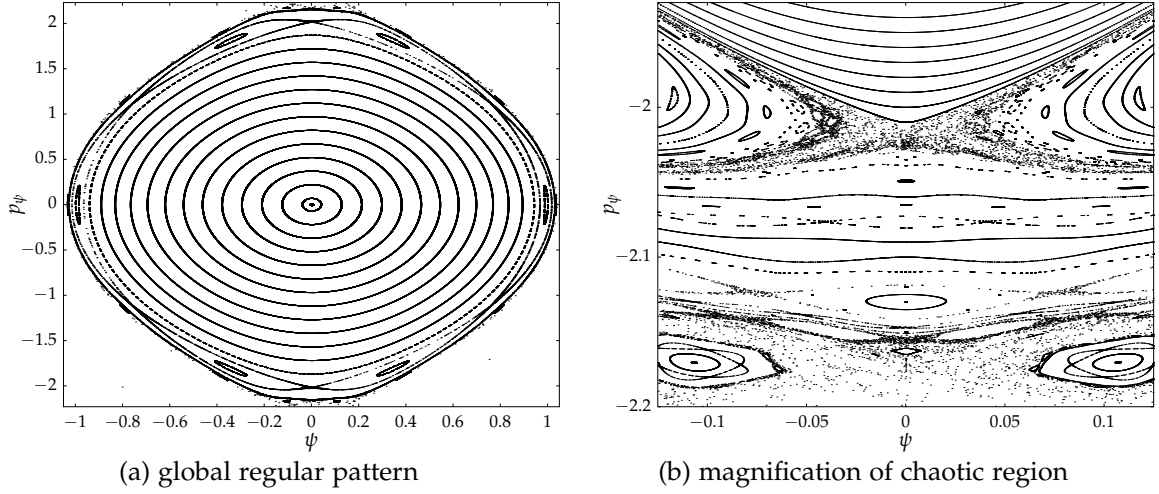

  \begin{center}
    \begin{subfigure}[b]{0.47\textwidth}
      \label{fig:angle25a}
      \resizebox{75mm}{!}{\input{aa.tex}}
      \caption{global regular pattern \label{fig:aa}}
    \end{subfigure}\quad\quad
    \begin{subfigure}[b]{0.47\textwidth}
      \resizebox{75mm}{!}{\input{aazoom.tex}}
      \caption{magnification of chaotic region \label{fig:aazoom}}
    \end{subfigure}
    \caption{\small The Poincar\'e cross section for the
      Hamiltonian~\eqref{eq:7} on energy level $e=50$, for $m=-3$ and
      $\kappa=1$. The cross-section plane is $r=\pi/2$ with $p_r>0$}
  \end{center}
\end{figure*}
For this potential we have $\lambda=(m+1)/m$. Comparing this value
with the forms of $\lambda$ given in the cases $1,2,3,4$ of the
integrability table~\eqref{eq:tab}, we obtain two values: $m=-1$, and
$m=-3$. In order to check if for these values of $m$ system is
integrable we made Poincar\'e cross-sections.  For $m=-1$,
cross-section in Figure~\ref{fig:km1} made for $\kappa=1$ clearly
shows that the system is not integrable.  For the case $m=-3$ the
answer was not obvious.  We repeated our investigations for different
choices of energy with the same results, namely the system behaves as
a regular one, see in Fig.~\ref{fig:km3glob} made for $\kappa=1$. Only
a big magnification of the region around an unstable periodic solution
shows a small region with chaotic behaviour, see in Fig.~\ref{fig:km3mag}.

It appears that chaotic region for this system can be detected much
easier in the complex part of the phase space.  Notice that after the
change of variables $\varphi=\rmi\psi,\ p_\varphi=-\rmi p_\psi$, the
Hamiltonian takes the form   
\begin{equation}
  \label{eq:7} H=\frac{1}{2}\left(p_r^2-\dfrac{p_{\psi}^2}{\sk(r)^2}\right)+\sk^m(r)\cos\psi.
\end{equation}

Now non empty constant real energy levels of Hamiltonian contain
points $(r,\psi,p_r,p_{\psi})\in\R^4$ which are not accessible in
original variables.  A global cross-section for
Hamiltonian~\eqref{eq:7} is presented in Fig.~\ref{fig:aa}. It shows
a small chaotic area in the vicinity of a certain unstable periodic
solution, see magnification of this region presented in
Figure~\ref{fig:aazoom}. It is a few orders bigger in comparison with
the chaotic region in Fig.~\ref{fig:km3mag}. We observed earlier this
phenomenon for the  Gross-Neveu systems in
\cite{Maciejewski:05::c}.

We can conclude that the potential~\eqref{eq:potential example 3} with
$\kappa\neq 0$ and for $m=-1,-3$, satisfying the necessary
integrability conditions, is not integrable.  However, in the case of
flat space when $\kappa=0$, and for $m=-3$ the system is integrable
with a first integral quartic in momenta of the following form
\begin{equation}
  I=p_\varphi^4+2r^{-1}p_\varphi^2\cosh\varphi+
  2p_rp_\varphi\sinh\varphi-r^{-2}\sinh^2\varphi.
\end{equation}
This case admits a generalisation. We can take
\begin{equation}
  \label{eq:function for 4 integral}
  U(\varphi)=Ae^{\varphi}+Be^{-\varphi},
\end{equation}
where $A$ and $B$ are  arbitrary constants, and then the quartic
first integral has the form
\begin{equation}
  I=p_\varphi^4+2r^{-1}U(\varphi)p_\varphi^2
  +2U'(\varphi)p_rp_\varphi -r^{-2}\left(U'(\varphi)\right)^2.
\end{equation}

Our last example is system~\eqref{eq:homo} with $m=2$ and with the
potential
\begin{equation}
  \label{eq:9}
  V(r,\varphi)=\sk^2(r)U(\varphi), \qquad U(\varphi)=c_1\cos(2\varphi) + c_2\sin(2\varphi),
\end{equation} 
where $c_1$ and $c_2$ are arbitrary constants. In this case system is
integrable with the following first integral
\begin{equation}
  \label{eq:10}
  I = \left[ p_r^2 - \left(p_{\varphi} \frac{\ck(r)}{\sk(r)} \right)^2\right]U(\varphi) + p_rp_{\varphi} \frac{\ck(r)}{\sk(r)}U'(\varphi) +2(c_1^2+c_2^2)\sk(r)^2.
\end{equation}

\section*{Acknowledgement}
The work  has been supported by grant No. DEC-2013/09/B/ST1/04130 of National Science Centre of Poland. 
\appendix
\section{Gauss hypergeometric equation} 
\label{sec:hyper}
The  Riemann $P$ equation  is the most general
second order differential equation with three regular singularities  \cite{Whittaker:35::}.
If we place using homography these singularities at $z\in\{0,1,\infty\}$,
then it has the canonical form
\begin{equation}
\label{eq:riemann}
\begin{split} 
\dfrac{\mathrm{d}^2\eta}{\mathrm{d}z^2}&+\left(\dfrac{1-\alpha-\alpha'}{z}+ 
\dfrac{1-\gamma-\gamma'}{z-1}\right)\dfrac{\mathrm{d}\eta}{\mathrm{d}z}+ 
\left(\dfrac{\alpha\alpha'}{z^2}+\dfrac{\gamma\gamma'}{(z-1)^2}+ 
\dfrac{\beta\beta'-\alpha\alpha'-\gamma\gamma'}{z(z-1)}\right)\eta=0, 
\end{split} 
\end{equation}
where $(\alpha,\alpha')$, $(\gamma,\gamma')$ and $(\beta,\beta')$ are the
exponents at the respective singular points. These exponents satisfy
the Fuchs relation
\[
\alpha+\alpha'+\gamma+\gamma'+\beta+\beta'=1.
\]
We denote the differences of exponents by
\[
\rho=\alpha-\alpha',\qquad \tau=\gamma-\gamma',\qquad \sigma=\beta-\beta'.
\]

Necessary and sufficient conditions for solvability of the identity
component of the differential Galois group of \eqref{eq:riemann}  are
given by the following theorem due to Kimura  \cite{Kimura:69::}.
\begin{table}[h]
 \begin{tabular}{lllll}
    \toprule
        \text{1}&$1/2+r$&$1/2+s$&arbitrary complex number&\\
        \text{2}&$1/2+r$&$1/3+s$&$1/3+p$&\\
        \text{3}&$2/3+r$&$1/3+s$&$1/3+p$&$r+s+p$ even\\
        \text{4}&$1/2+r$&$1/3+s$&$1/4+p$&\\
        \text{5}&$2/3+r$&$1/4+s$&$1/4+p$&$r+s+p$ even\\
        \text{6}&$1/2+r$&$1/3+s$&$1/5+p$&\\
        \text{7}&$2/5+r$&$1/3+s$&$1/3+p$&$r+s+p$ even\\
        \text{8}&$2/3+r$&$1/5+s$&$1/5+p$&$r+s+p$ even\\
        \text{9}&$1/2+r$&$2/5+s$&$1/5+p$&\\
        \text{10}&$3/5+r$&$1/3+s$&$1/5+p$&$r+s+p$ even\\
        \text{11}&$2/5+r$&$2/5+s$&$2/5+p$&$r+s+p$ even\\
        \text{12}&$2/3+r$&$1/3+s$&$1/5+p$&$r+s+p$ even\\
        \text{13}&$4/5+r$&$1/5+s$&$1/5+q$&$r+s+p$ even\\
        \text{14}&$1/2+r$&$2/5+s$&$1/3+p$& \\
        \text{15}&$3/5+r$&$2/5+s$&$1/3+p$&$r+s+p$ even\\
  \bottomrule
  \end{tabular}
   \caption{Schwarz's table. Here $r,s,p\in\Z$ \label{tab:sch_app}}
\end{table}
\begin{theorem} 
\label{th:Kimura}
  The identity component of the differential Galois group of the Riemann $P$ equation \eqref{eq:riemann} is solvable iff
\begin{description} 
\item[A.] at least one of the four numbers $\rho+\sigma+\tau$,
  $-\rho+\sigma+\tau$, $\rho+\sigma-\tau$, $\rho-\sigma+\tau$ is an odd
  integer, or
\item[B.] the numbers $\rho$ or $-\rho$  and
  $\sigma$ or $-\sigma$ and $\tau$ or $-\tau$ belong (in an arbitrary order) to some of
  appropriate fifteen families forming the so-called Schwarz's Table, see Table~\ref{tab:sch_app}.
\end{description}
\end{theorem}

\bibliographystyle{klunamed}
\newcommand{\noopsort}[1]{}\def\polhk#1{\setbox0=\hbox{#1}{\ooalign{\hidewidth
  \lower1.5ex\hbox{`}\hidewidth\crcr\unhbox0}}} \def\cprime{$'$}
  \def\cydot{\leavevmode\raise.4ex\hbox{.}} \def\cprime{$'$} \def\cprime{$'$}
  \def\cprime{$'$} \def\cprime{$'$} \def\cprime{$'$} \def\cprime{$'$}
  \def\cprime{$'$} \def\cprime{$'$} \def\cprime{$'$} \def\cprime{$'$}

\end{document}